\documentclass[prb,floatfix,twocolumn,amsmath,amssymb]{revtex4}
\usepackage{graphicx}
\usepackage{dcolumn}
\usepackage{bm}
\begin{document}

\title{Stability of RVB hole stripes in high-temperature superconductors}

\author{Manuela Capello, $^{1}$ Marcin Raczkowski, $^{2}$ Didier Poilblanc $^{1}$}
\affiliation{
$^{1}$  Laboratoire de Physique Th\'eorique UMR 5152, CNRS and 
Universit\'e Paul Sabatier, F-31062 Toulouse, France\\ 
$^{2}$ Marian Smoluchowski Institute of Physics,
Jagellonian University, Reymonta 4, PL-30059 Krak\'ow, Poland }

\begin{abstract}
Indications of density-wave states in underdoped cuprates, coming
from recent STM 
(scanning tunneling microscopy) and Hall-resistance measurements, 
have raised new concerns whether stripes could be stabilized in
the superconducting phase of cuprate materials, even
in the absence of antiferromagnetism.
Here, we investigate this issue using state-of-the-art quantum Monte Carlo 
calculations of a $t-J$ model.
In particular we consider the stability of
unidirectional hole domains in a modulated 
superconducting background, by taking into account the effect of
tetragonal-lattice distortions, next-nearest neighbor hopping and
long-range Coulomb repulsion. 
\end{abstract}
\date{\today}
\maketitle
\section{Introduction}
The physics of high-temperature superconductors (HTSC)
is characterized by the presence of several competing orders.
Besides antiferromagnetism and superconductivity, it is now widely accepted
that the underdoped region of the HTSC phase diagram shows 
a pseudogap phase~\cite{arpes1,arpes2}
and occasionally the presence of spatially-ordered
states~\cite{tranquada,kivelson} at special dopings, 
characterized by charge domain walls separated by antiferromagnetic
regions.
Recently, accurate STM experiments~\cite{kohsaka}
have revealed modulations of the local density of states
in the form of unidirectional
domains for two different underdoped
HTSC materials, Ca$_{2-\delta}$Na$_\delta$CuO$_2$Cl$_2$ and 
Bi$_2$Sr$_2$CaCu$_2$O$_{8+\delta}$,
where spatial ordering was not detected before.
These findings highlight the importance of inhomogeneous states
as a general feature of HTSC, and
extend the idea of quasi-one dimensional order beyond the original
spin-stripe picture~\cite{stripe}, with superconductivity
coexisting with charge modulation, in the absence of antiferromagnetism. 
Moreover, recent measurements of Hall resistance~\cite{leboeuf} on underdoped 
YBa$_2$Cu$_3$O$_\delta$ and YBa$_2$Cu$_4$O$_8$ give evidence of
a reconstruction of the Fermi surface caused by the onset of a density-wave 
phase in the large-field induced normal state, 
indicating that spatial symmetry breaking is a common characteristic
of underdoped cuprates.

Theoretically, it is challenging to 
characterize a modulated state that could mimic
the experimental findings, and to identify
the key microscopic parameters 
that could induce this kind of non-uniform superconducting phases.
Indeed, besides the strong, local repulsion among electrons, that leads to
antiferromagnetism,
the low-energy physics of cuprates is often determined by additional
hopping integrals among next-nearest neighbor copper atoms,
and by a long-range Coulomb repulsion among electrons.
These terms have been studied in the past for different
inhomogeneous states and were found to stabilize modulated 
superstructures~\cite{weber,ogata}. 
Also of relevance, is the role played by
structural instabilities, a feature undoubtedly present in
many HTSC compounds, in stabilizing modulated superstructures.
Indeed, HTSC are often characterized by a
low-temperature tetragonal phase~\cite{buchner}, produced by a tilt
of the oxygen octahedra that leads to a different electronic
hopping and exchange integral along the two planar directions.
Previous calculations found that
spatial anisotropy could
stabilize antiferromagnetic-stripe phases~\cite{kampf,becca}.
Therefore, in view of what is found in experiments, it is 
important to understand if a similar effect occurs for {\em superconducting}
modulated states.

In this paper, we study the competition
among homogeneous and non-uniform superconducting states, 
by taking into account the presence of spatial lattice distortions,
next-nearest neighbors hopping terms and long-range Coulomb repulsion.
As a prototype model for cuprate superconductors, 
we investigate the $t-J$ model~\cite{rice}
in two dimensions, 
by using state-of-the-art Variational Monte Carlo techniques (supplemented
by mean-field approaches).
Following Anderson idea~\cite{anderson},
the variational approach, based on the resonating-valence-bond
(RVB) wavefunction, has successfully described
most of the features of HTSC~\cite{gazza,spanu}.
However, to tackle
spatially-inhomogeneous superconducting states 
like those found by experiments,
more involved calculations, including the possibility of
local modulations in the superconducting state,
are required. 
Here, we accomplish this task and show that, generically, 
half-filled hole stripes can form easily in a RVB superconductor at quite 
small energy cost. Interestingly enough, lattice distortion
and long-range Coulomb repulsion
are proved to play a key role.  

This paper is organized as follows. In Section~\ref{secmodel}
we introduce the model and the variational approach, presenting
the two families of modulated wavefunctions that we study, the
$\pi$-phase shift and inphase domain RVB wavefunctions.
In Section~\ref{sectj} we characterize in detail the 
new inphase domain RVB state, introduced in this paper, 
for the $t-J$ model.
In Section~\ref{sectetragonal},~\ref{sectp} and~\ref{secV}
we show the effect of lattice distortion, next-nearest neighbor hopping and 
long-range Coulomb repulsion, respectively, 
on the stabilization of modulated superconducting states.
Finally, in Section~\ref{secsummary} we draw our conclusions.
\section{MODEL and WAVEFUNCTIONS\label{secmodel}}
The $t-J$ Hamiltonian is defined, using standard notations, as~\cite{rice}:
\begin{equation}\label{Htj}
H_{t-J}=-t\sum_{\langle i,j\rangle \sigma} 
\alpha_{ij} (\tilde c^\dagger_{i\sigma} \tilde c_{j\sigma}+ h.c.)+  
J\sum_{\langle i,j\rangle} \alpha^2_{ij} {\bf S}_i\cdot {\bf S}_j
\end{equation}
where $\tilde c^\dagger_{i\sigma}=(1-n_{i\, -\sigma})c^\dagger_{i\sigma}$
acts on the reduced Hilbert space with no double occupancies.
For non-distorted lattices we take $\alpha_{ij}=\alpha_x=\alpha_y=1$
for all nearest neighbors.
Instead, tetragonal distortion  is set by taking different hopping
and exchange parameters along the $x$ and $y$ 
direction, according to the position of the tilt axis.
This implies 
$\alpha_x\ne\alpha_y$, with $\alpha_{x} < 1$ ($\alpha_{y} < 1$) and 
$\alpha_{y} =1$ ($\alpha_{x} =1$) respectively 
for the tilt axis along the $y$ ($x$) direction.
Here, we mainly focus
on the physically relevant case $t/J=3$ (if not specified otherwise)
at doping $\delta=1/8$.

In the following, we make use of the Variational Monte Carlo techniques
(VMC),
using clusters
of $N = 8\times  8$, 128 (45 degree $8^2+8^2$ tilted lattice) and 
$16 \times 16$ sites  with
periodic-boundary conditions.
Moreover, to have access to larger sizes that can approach
the thermodynamic limit, we compare our variational results with 
predictions coming from the renormalized mean-field theory 
(RMFT)~\cite{zhang}, on a $128 \times 128$ cluster, 
using unit cell translation symmetry~\cite{Rac06} (for more
details on the RMFT results, see Ref.~\cite{Rac08}).
In the variational procedure, the
RVB state is derived from the BCS mean-field state
by incorporating the effect of correlation 
via the so-called Gutzwiller projector ${\cal{P}}_g$, that 
can be treated exactly
within the quantum Monte Carlo scheme.
Indeed, given the mean-field Hamiltonian:
\begin{eqnarray}\label{eqHmf}
H_{MF}=\sum_{i,j\sigma} 
(\chi_{ij} c^\dagger_{i\sigma} c_{j\sigma}+ h.c.)+ \nonumber \\ 
+\sum_{\langle i,j\rangle} 
(\Delta_{ij} c^\dagger_{i\uparrow} c^\dagger_{j\downarrow}+ h.c.)
+ \mu\sum_{i\sigma} n_{i\sigma}  
\end{eqnarray}
we construct the variational
state by applying the Gutzwiller projector to
the ground state $|D\rangle$ of Eq.~(\ref{eqHmf}):
\begin{equation}\label{eqGutz}
|\Psi_{G}\rangle={\cal{P}}_g |D\rangle=
\prod_i(1-n_{i\uparrow}n_{i\downarrow})|D\rangle
\end{equation}
where $n_{i\sigma}$ counts the number of electrons of spin $\sigma$
on site $i$.
The terms $\chi_{ij}$, $\Delta_{ij}$
and $ \mu$ in Eq.~(\ref{eqHmf}) correspond to a set of parameters that are
optimized in order to minimize the variational energy, according to
the stochastic minimization algorithm~\cite{sorella}.
In the simple case of the uniform RVB state, the independent parameters
reduce to $\mu$ and $\Delta_{ij}=\pm \Delta$ for nearest neighbors, 
with the sign following the $d$-wave symmetry.
Moreover, another competing homogeneous state, the staggered-flux 
phase~\cite{sfp} (SFP),
is obtained by allowing the hopping parameters $\chi_{ij}$ to be  
complex, leading to staggered currents circulating in
opposite directions in the neighboring
plaquettes.

Here, we study possible instabilities of the RVB state
towards spatial modulations, allowing 
the RVB bonds $\Delta_{ij}$  to become inhomogeneous.
In particular, it is known from experiments that
one-dimensional hole-rich regions are generally spaced by
$1/(2\delta)$, implying half-filled hole domains
with an average of one hole every $1/\delta$ sites
in the direction where translational symmetry is broken.
At a doping of $\delta=1/8$,
this corresponds to a periodicity of four lattice spacings $4a$.
Remarkably, we found such states, where 
hole-rich stripes coexist with modulated superconducting domains,
whose energy per site is only of a very small fraction of $J$ higher
than its uniform counterpart.
The hierarchy of phases is therefore expected to be
very sensitive to the details of the microscopic Hamiltonian.

\begin{figure}
\includegraphics[width=\columnwidth]{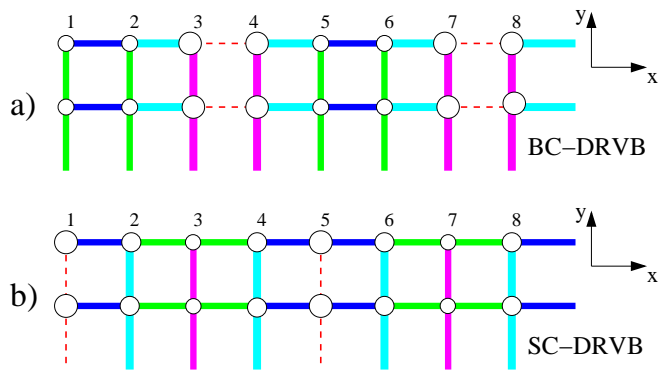}
\caption{\label{fig1}
(Color online)
Cartoon of the
(a) bond-centered and (b) site-centered domain RVB (DRVB) stripes.
Non-equivalent bonds have different colors.
Thicker (larger) bonds (circles) have larger pairing (hole)
densities.
Dashed lines indicate bonds where ${\Delta_{ij}}$ is set to zero.
}
\end{figure}
In a previous paper~\cite{marcin}, 
we considered a variational 
domain-RVB state
that contains a $\pi$-phase shift in the superconducting order parameter
between regions of four-lattice spacings width ($\pi$DRVB). 
The phase shifts create
domain walls with vanishing pairing amplitude
and a consequent concentration of holes.
The $\pi$DRVB states turned out to be competitive
with respect to the homogeneous RVB state, and could explain the
copper-oxygen layer decoupling found in La$_{2-x}$Ba$_x$CuO$_4$,
accompanied by a depression of $T_c$ at 1/8 doping~\cite{li,fradkin}.
Here, we complete the characterization of $\pi$DRVB states, 
by considering the effect of lattice distortion, nearest-neighbor
hopping term, and long-range Coulomb repulsion. 
Moreover, since the presence of the antiphase domains implies a certain
amount of energy,
in the following we investigate an alternative strategy to tackle the problem,
and propose another possible candidate for superconducting stripes, 
where domains are inphase: the domain RVB state (DRVB).
Starting from the RVB state, we investigate the
energy cost associated to the introduction of line defects
in the uniform state, in the form of unidirectional hole domains. 
This is done by imposing a vanishing pairing amplitude $\Delta_{ij}=0$ along
one direction, with a periodicity that, at doping 1/8, corresponds to 4$a$.
In Figure~\ref{fig1} we show the two possible modulations of the inphase
DRVB state derived from the above arguments.
In Figure~\ref{fig1}(a) stripes
are bond-centered (BC), with the RVB bonds characterized by
a periodicity of $4a$ along the $x$ direction. 
The pairing field $\Delta_{ij}$ 
associated to one type of horizontal bonds is set to zero.
Alternatively, 
in Figure~\ref{fig1}(b) the stripes are site-centered (SC),
where the lines of vertical bonds with $\Delta_{ij}=0$ 
are separated by $4a$
along the $x$ direction.

Besides the $\pi$-shift domain RVB stripes, which
could give a natural explanation 
to the decoupling among superconducting-layers,
this second class of superconducting hole stripes, the DRVB states,
could be related to the STM
experimental findings of inhomogeneous superconducting states.
\section{DRVB stripes in the $t-J$ model}
\label{sectj}
In this section we characterize the new inphase domain RVB states
for the $t-J$ model, showing that, among the
superconducting stripe states~\cite{marcin,vojta},
they constitute a valid candidate for
a strongly-competing, inhomogeneous state.
We start by considering the variational energies of the 
different stripe states, together with the corresponding 
RMFT values 
in Table~\ref{tabenergy}. 
\begin{table}
\caption{\label{tabenergy} VMC and RMFT energy per site (in units of $t$)
for different projected wavefunctions for the
$t-J$ model at doping 1/8.}
\begin{tabular}{ccc}
\hline \hline
$|D\rangle$ & $E_{VMC} \; [\, t\, ]$ &$E_{RMFT} \; [\, t\, ]$ \\
\hline \hline
RVB                           & -0.45564(3) & $-$0.4549(1) \\
SFP                           & -0.44630(3) & $-$0.4286(1) \\
$\pi$DRVB~[Ref.\cite{marcin}] & -0.44529(3) & $-$0.4413(1) \\
BC-DRVB                      & -0.45490(3) & $-$0.4511(1) \\
SC-DRVB                      & -0.45525(3) & $-$0.4507(1) \\
\hline \hline
\end{tabular}
\end{table}
Both approaches show that, with respect to the 
SFP and the $\pi$DRVB~\cite{marcin}
states, the DRVB states are one order of magnitude
closer in energy to the homogeneous RVB phase. 
Note that the site-centered variational energy approaches the RVB state
very closely. However, 
in both cases the small energy difference 
indicates both DRVB states
 as very promising candidates for a 
competing inhomogeneous state. 
Remarkably, within the DRVB geometry,
the suppression of some of the RVB bonds 
does not imply significant energy costs.
Notice that, at the variational level, the optimal wavefunctions preserve 
the $d$-wave symmetry of the pairing field 
in the regions where $\Delta_{ij}\ne 0$, with a small modulation
that depends upon the type of bond (see Figure~\ref{fig1}). 

\begin{figure}
\includegraphics[width=\columnwidth]{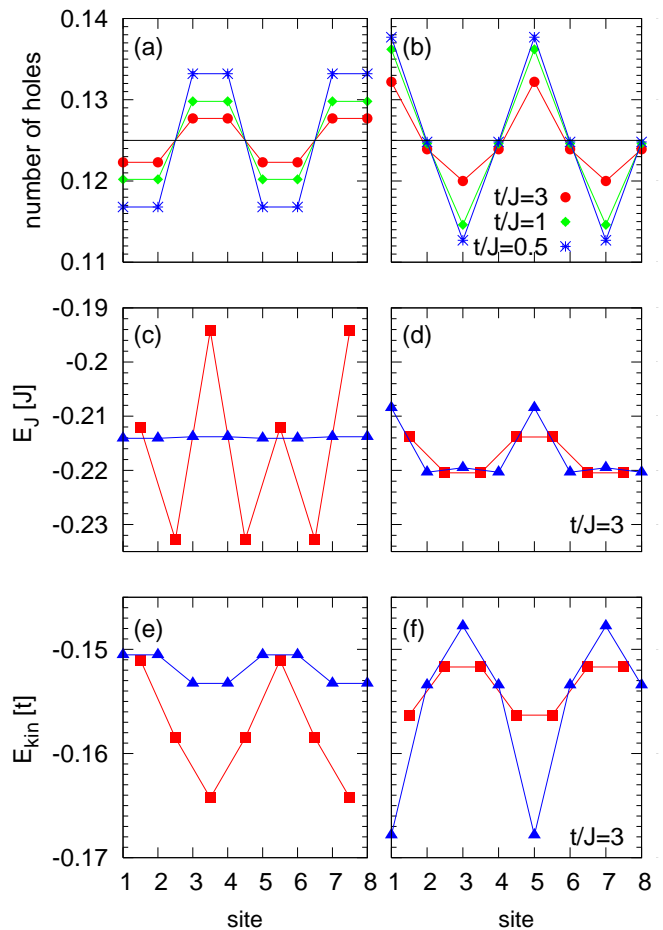}
\caption{\label{plot-hole-EJ-Ek}
(Color online)
Top panel:
Hole distribution in DRVB states
for different values of $t/J$ as shown in graph.
Middle and bottom panel: Magnetic (in units of $J$)
and kinetic (in units of $t$) bond energy  within the unitary cell
of Figure~\ref{fig1} for $t/J=3$.
Squares indicate energies associated to horizontal bonds
(located between two sites), triangles to vertical bonds.
Left panels refer to the
BC-DRVB state and right panels to the SC-DRVB state. All quantities are
calculated for a cluster of
16x16 sites.}
\end{figure}

\begin{figure}
\includegraphics[width=\columnwidth]{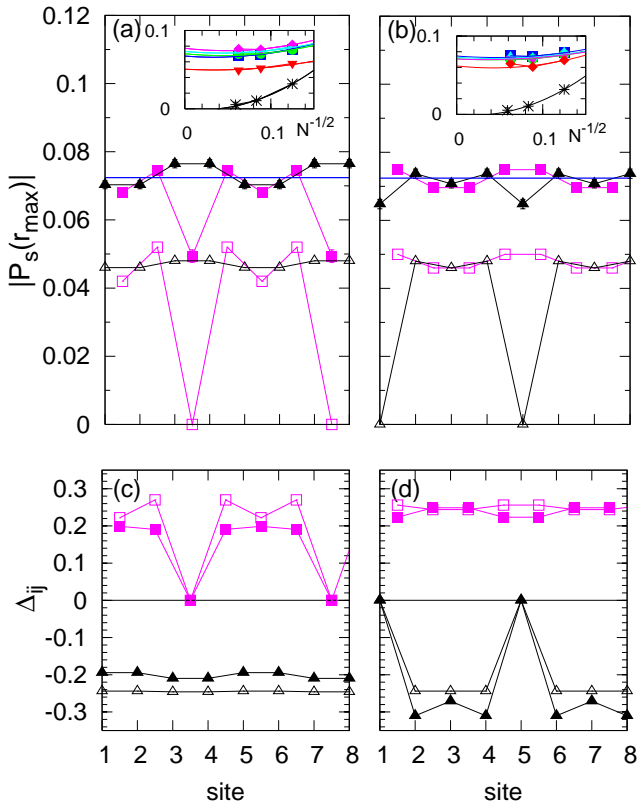}
\caption{\label{pair+paramBC}
(Color online)
Top panel: Pairing order parameter (filled points for VMC,
empty points for RMFT) for the BC-DRVB (left) and SC-DRVB (right)
states along the 8 independent sites of Figure~\ref{fig1}.
Squares correspond to horizontal bonds (located between two sites),
triangles to vertical bonds.
The blue line corresponds to the pairing order parameter
in the homogeneous RVB state.
Inset: size scaling of the pairing order parameter as a function of the
inverse-squared number of sites.
Colors refer to the corresponding colored bonds
of Figure~\ref{fig1}. Stars correspond to the projected Fermi sea,
for comparison.
Bottom panel: corresponding pairing variational parameters
for the different bonds.
}
\end{figure}

To prove that the non-homogeneous DRVB wavefunctions that we have constructed
indeed correspond to spatially modulated states, we show
in Figure~\ref{plot-hole-EJ-Ek}
typical profiles in the bond- and site-centered states.
It turns out that the hole distribution along the different sites of the unitary
cell is non-uniform, the
hole density being larger along the  regions where $\Delta_{ij}=0$
(notice that the magnitude of the hole modulation increases with $J/t$).
Moreover, 
the magnetic and kinetic energies follow the same modulation than the charge. 
Indeed, along the $\Delta_{ij}=0$ bonds 
 the kinetic energy is enhanced, while the magnetic energy is suppressed,
as expected.
Note that the bonds presenting the largest modulation
are the horizontal (vertical) ones for the BC (SC) 
wavefunction, i.e., the bonds along the direction where some
of the pairing fields are suppressed. 

Finally, to characterize 
the superconducting properties of our variational states,
we calculate the singlet pairing correlations:
\begin{equation}\label{eqpairing}
P_s^2(r) 
= 
\frac{\langle \Psi_G| \tilde\Delta^\dag_{s+r} \tilde\Delta_{s} |\Psi_G \rangle}
{\langle \Psi_G|\Psi_G \rangle},
\end{equation}
where
$\tilde\Delta^\dag_{s} = c^\dag_{s,\uparrow} c^\dag_{s+{\hat a},\downarrow} -
c^\dag_{s,\downarrow} c^\dag_{s+{\hat a},\uparrow}$
creates a singlet pair of electrons among nearest neighbors 
for each independent site $s$ of the unitary cell in Figure~\ref{fig1} and 
${\hat a}$
is the unit vector, that specifies the bond direction (along $x$ or $y$).
We extract the pairing order parameter by taking the square root 
$P_s(r_{max})$ 
of the pairing correlation function (\ref{eqpairing})
at the maximum distance $r_{max}$ 
for different cluster sizes, for all independent sites. 
Notice that the VMC and RMFT profiles are in agreement.
However, within VMC, we find, for all different bonds, 
that the pairing order parameter is
finite, see Figure~\ref{pair+paramBC}. 
Remarkably, this is the case also for the bond where the variational parameter
$\Delta_{ij}=0$, 
signaling that the Gutzwiller approximation  is not accurate
in this case. The resulting picture consists of two different types of bonds,
some associated to the hole-rich region, and others
 where the superconducting order parameter, although slightly enhanced
or depressed locally, is very close the one found for the 
homogeneous RVB state.  
This feature, together with the non-uniform charge distribution,
has strong similarities with the picture which could be derived from
the STM experimental findings of Ref.~\cite{kohsaka}.

\section{Lattice distortion}
\label{sectetragonal}
\begin{figure}
\includegraphics[width=\columnwidth]{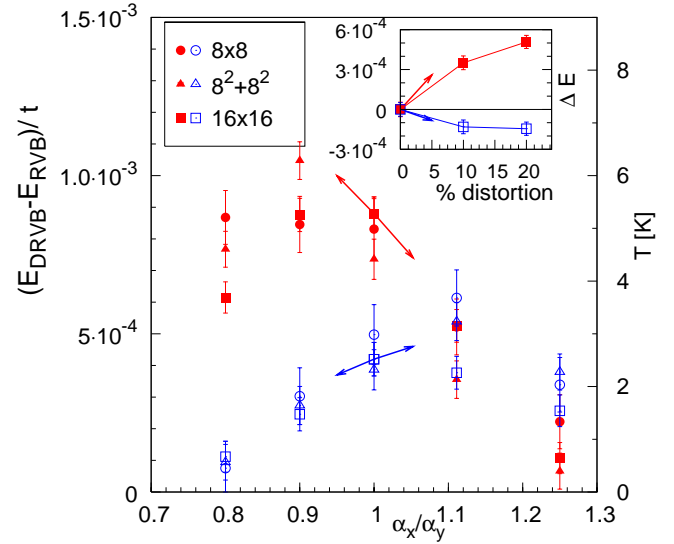}
\caption{\label{energy-J-tx}
(Color online)
Energy difference between the  BC-DRVB (SC-DRVB) and
the homogeneous RVB state~\cite{note2}
as a function of the lattice distortion $\alpha_x/\alpha_y$,
shown by filled (empty) symbols.
Charge stripes are arranged along the $y$ direction in all cases.
Energies (per site) are reported in units of $t$, for the case $t/J=3$.
The right axis is a temperature scale, assuming $J=2000K$.
The points considered correspond to distortions of 10$\%$
and 20$\%$ in the two directions.
Inset: Energy difference among the two possible stripe arrangements,
parallel and perpendicular to the tilt axis, for BC and SC
wavefunctions. Arrows correspond to the predictions given by the
Hellman-Feynman theorem for the largest system.
}
\end{figure}

Considering the $t-J$ Hamiltonian
of Eq.~(\ref{Htj}), lattice distortion is set by imposing
two different $\alpha_{ij}$ along the $x$ and $y$ direction. 
We find that setting $\alpha_x\ne\alpha_y$ further increases the stabilization
of the non-homogeneous states.
This feature, that was already observed in the context of
spin-stripes~\cite{kampf,becca}, extends its validity in the quite
different case of superconducting-stripe states.
Considering the case of DRVB states,
we find that the bond-centered stripes gain
a remarkable amount of energy upon distortion, see Figure~\ref{energy-J-tx}.
In particular,
we find that the superconducting inphase 
domain stripes are stabilized~\cite{note}
when the $\Delta_{ij}=0$ bonds lie along the direction of the
tilt axis,
where the hopping and exchange parameters are the largest.
Assuming by convention the hole-stripe fixed along the $y$ direction,
we indeed found the bond-centered hole
stripes favored for $\alpha_x>\alpha_y$ (tilt axis along $x$, with
the stripe perpendicular to it)
and the site-centered hole stripes favored for $\alpha_x<\alpha_y$
(tilt axis along $y$, i.e., parallel to the stripes, similarly to what
found for antiferromagnetic stripes in Ref.~\cite{kampf}).
The stability of the two possible stripe arrangements is evident
in the inset of Figure~\ref{energy-J-tx}, where
$\Delta E$, i.e., the energy of the stripe
parallel to the tilt axis minus the energy of the stripe
perpendicular to it, shows a different behavior for BC and SC stripes.
Note that all data are in agreement with calculations using
the Hellman-Feynman theorem (providing directly the slopes in the limit
$\alpha_{x (y)}\to 1$),
valid for small deformations,
which gives $\Delta E/t=0.00520 \Delta t/t$
for bond-centered stripes and $\Delta E/t=-0.00167 \Delta t/t$
for site-centered stripes.
Remarkably, the RMFT approach also confirms that these
superconducting modulated states are
stabilized by tetragonal distortion (not shown).
\begin{figure}
\includegraphics[width=\columnwidth]{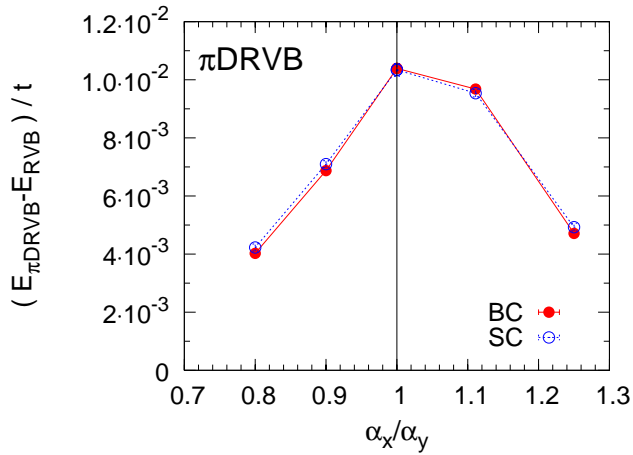}
\caption{\label{tetragonalpi}
(Color online)
Energy difference between the  BC-$\pi$DRVB (SC-$\pi$DRVB) and
the homogeneous RVB state~\cite{note2}
as a function of the lattice distortion $\alpha_x/\alpha_y$,
shown by filled (empty) symbols.
Charge stripes are arranged along the $y$ direction in all cases.
Energies (per site) are reported in units of $t$.
The points considered correspond to distortions of 10$\%$
and 20$\%$ in the two directions.
}
\end{figure}
Instead, in the case of $\pi$-shift domain stripes, which involves
higher energy scales, there is no
appreciable difference in the behavior of the
bond- and site-centered geometries with respect to the distortion axis,
see Figure~\ref{tetragonalpi}.
Indeed, lattice deformation lowers the $\pi$DRVB energy in both cases
with respect to the uniform RVB state,
with a slightly more favorable contribution when the distortion
axis lies on the $y$ direction, i.e. parallel to the stripes.

\section{Role of $t'$}
\label{sectp}
\begin{figure}
\includegraphics[width=\columnwidth]{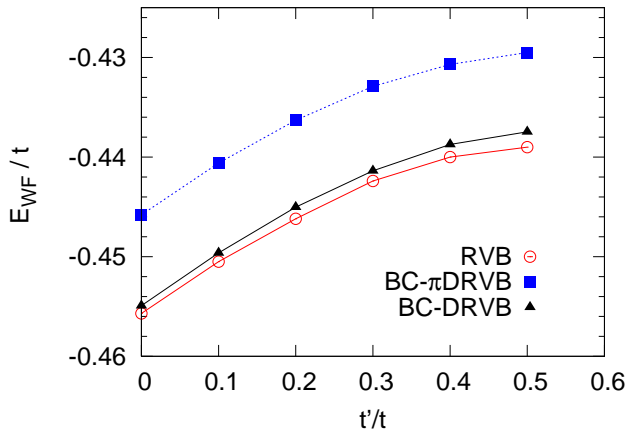}
\caption{\label{fig5}
(Color online) $t-t'-J$ model:
Variational energy (in units of $t$) vs. $t'/t$ for the uniform RVB
(circles),
the BC-$\pi$DRVB (squares) and the BC-DRVB (triangles) wavefunctions.}
\end{figure}

Since the energies of the stripe states are very close
to the homogeneous RVB state, 
some further contribution to the microscopic Hamiltonian
could finally stabilize the bond-centered 
hole-stripe states as found in experiments. Here, we consider
an additional hopping integral among next-nearest neighbor copper atoms, 
which is relevant, in many cuprates, for
a correct description of their Fermi surface.
The $t-t'-J$ Hamiltonian is obtained from the $t-J$ model of Eq.
(\ref{Htj}) by adding:
\begin{eqnarray}
H_{t-t'-J}=H_{tJ}+
t'\sum_{\langle\langle i,j\rangle \rangle\, \sigma} 
(c^\dagger_{i\sigma} c_{j\sigma}+ h.c.)
\end{eqnarray}
where 
$\langle\langle i,j\rangle \rangle$ denotes next-nearest neighbors.
Here, we take $t'>0$ and $t/J=3$, being relevant for hole-doped cuprates.
In the presence of $t'$, a further variational hopping parameter 
$\chi'_{ij}$
among next-nearest neighbors is added to the mean-field Hamiltonian
of Eq.~(\ref{eqHmf}).
Considering the uniform RVB state, 
$\chi'_{ij}$ is homogeneous for all sites.
Instead, in the case of stripe states, four different 
$\chi'_{ij}$ are optimized, according to the symmetries of the
$\pi$DRVB and DRVB unitary cells.
In Figure~\ref{fig5} we compare the variational energies of the two 
candidate stripe states, the $\pi$-phase domain RVB stripe
and the inphase domain RVB stripe, with the energy of the homogeneous RVB state
for different $t'$. It turns out that the DRVB stripe is the most stable 
inhomogeneous state, but, in both cases, the effect of $t'$
is irrelevant for the stabilization of superconducting stripes.
\section{Role of non-local Coulomb repulsion}
\label{secV}
Finally, we consider the effect of a long-range Coulomb repulsion,
which could play an important role in the STM measurements.
To this purpose we consider the $t-J-V$ model:
\begin{eqnarray}
H_{t-J-V}= 
H_{tJ}+\sum_{ij} V_{|i-j|} (n_i-\bar n)(n_j- \bar n)
\end{eqnarray}
where  $\bar n$ is the average density of electrons.
The potential $V_{|i-j|}$ has the form of a screened Coulomb repulsion
$V_r=V \frac{\exp(-r/l_0)}{r}$, $r$ being the periodized
distance between different lattice sites.
In the following, we have fixed $l_0=4$
and considered the properties of the system as a function of $V$.
At the variational level, we find that, in the
presence of a non-local Coulomb repulsion,
a long-range Jastrow factor acting on the fully-projected mean-field state 
$|\Psi_{G}\rangle$ is needed:
\begin{equation}
|\Psi_{J}\rangle={\cal{P}}_{{{J}}} |\Psi_{G}\rangle
\end{equation}
where ${\cal{P}}_J=\exp\left[\frac{1}{2}\sum_{ij} v_{ij} n_i n_j\right]$, 
with $v_{ij}=v_{|i-j|}$
variational parameters which depend only on the distance $|i-j|$ and 
$|\Psi_{G}\rangle$ is defined in Eq.~(\ref{eqGutz}).
This term takes into account the two-body correlations along all possible
distances, and is necessary to get accurate wavefunctions in the presence 
of $V_r$.
\begin{figure}
\includegraphics[width=\columnwidth]{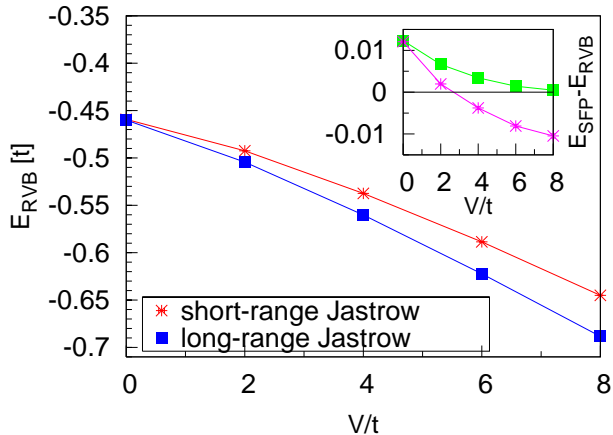}
\caption{\label{fig6}
(Color online)
Energy of the homogeneous RVB state as a function of the
Coulomb-repulsion strength $V/t$, in the presence of a short-range Jastrow factor
(stars) and a long range  Jastrow (squares).
Inset: Energy difference between the SFP and
the homogeneous RVB state as a function of the
Coulomb-repulsion strength $V/t$. Both wavefunctions are
optimized in the presence of a short-range (long-range) Jastrow factor,
the energy difference being represented by stars (squares).
}
\end{figure}
\begin{figure}
\includegraphics[width=\columnwidth]{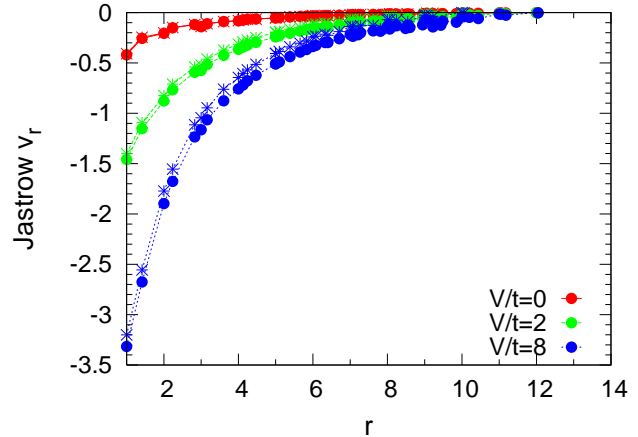}
\caption{\label{fig7}
(Color online)
 Jastrow parameter {\em vs.} distance for two tilted lattices on 242 (stars) and 338 sites (circles) and different values of $V/t$. }
\end{figure}
In Figure~\ref{fig6} we show the 
energy of the homogeneous RVB state, optimized with a short-range 
(i.e., up to nearest-neighbors) and a long-range Jastrow factor, respectively.
It turns out that, by increasing $V/t$, the long-range correlation term
plays a fundamental role in lowering the energy of the RVB variational state.
Therefore,
the addition of the long-range Jastrow term can strongly
influence the hierarchy of variational phases stabilized at large $V/t$.
Indeed, considering
relative stability of the SFP with respect to
 the RVB state (see inset of Figure~\ref{fig6}),
it turns out that, with a short-range Jastrow, the SFP is the 
lowest-energy state already for $V\simeq 3 t$, while the addition of a
long-range Jastrow stabilizes the RVB state up to very large
values of the Coulomb repulsion.
\begin{figure}
\includegraphics[width=\columnwidth]{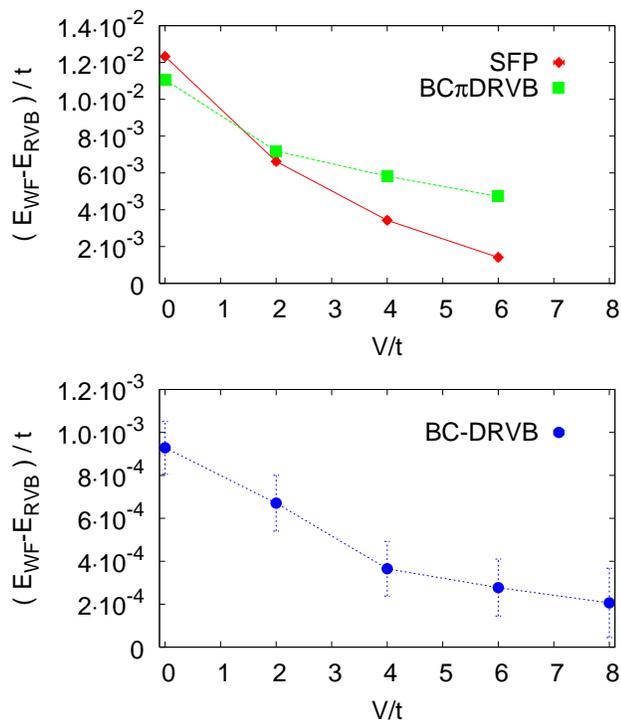}
\caption{\label{fig8}
(Color online)
Energy difference w.r.t. the uniform RVB state of
the SFP, $\pi$DRVB stripe state (top) and
the DRVB state (bottom), as a function of $V/t$. All wavefunctions
are optimized in the presence of a long-range Jastrow factor.}
\end{figure}
To better understand the role played by the Jastrow factor in the presence of 
$V_r$, we optimize the RVB wavefunction for
larger clusters. In Figure~\ref{fig7} we plot the resulting
Jastrow parameters in real
space. By increasing $V$, the correlation terms $v_r$ increase, as expected.
In particular, notice that the part at long distances is non-negligible
even for $r>l_0$. 
Since a uniform Jastrow can
promote the stability of uniform states and mask possible
instabilities towards modulated states, 
we optimize the Jastrow parameters up to the next-nearest neighbor distance
independently on the nonequivalent
bonds (i.e. here the $v_{ij}$ depend on $i$ and $j$). 
The resulting energies are shown 
in Figure~\ref{fig8} where, even though
the stripe energies approach the uniform RVB wavefunction by 
increasing $V/t$,
the homogeneous state still remains the most stable.
Notice that the presence of a long-range Jastrow factor stabilizes
the BC-$\pi$DRVB state w.r.t. the SFP, 
at $V=0$, contrarily to the values reported in Table~\ref{tabenergy}.
However, by increasing $V/t$ the SFP wavefunction gains energy and 
approaches closely the uniform RVB state. 
Moreover, considering the inphase DRVB state, it turns out that for finite $V/t$
the energy difference with respect to the RVB state
is very small, of the order of the error bars.
Unfortunately,
in the presence of a non-local Coulomb repulsion,
the main limit of this approach corresponds to the necessity of incorporating
long-range Jastrow terms in order to reach a good accuracy.
Since the energies into play are very small, the restriction of homogeneous
Jastrow terms at large distances, which governs 
the low-energy physics, could ultimately
favor the uniform RVB state and bias our variational outcome.
In other words, it might not be sufficient to assume
inhomogeneous Jastrow terms
at short distances.
However, these variational results 
suggest already that long-range Coulomb repulsion can be a
driving force for the stabilization of
modulated states.

\section{Summary and conclusions}
\label{secsummary}
In summary, we have shown that unidirectionally modulated superconducting states
are remarkably 
close in energy to the uniform RVB state.
This is the case both for $\pi$-phase shift domain RVB stripes
and for inphase domain RVB stripes.
Their properties reflect the recent STM observations, both concerning
the modulated charge and superconducting features.
Besides the lattice distortion, long-ranged Coulomb repulsion
further stabilizes the non-uniform superconducting stripe phases.
Instead, next-nearest neighbor hopping does not seem to play any role in the
hierarchy of phases studied here.
However, although the stripe energies can approach very closely
that of the homogeneous RVB state, so far the uniform solution
remains the most stable. 
In particular, we have found that a $\pi$-phase shift
costs more energy than the inphase counterpart having simple $\Delta=0$ domains.
Experiments suggest that the real stabilization of hole-stripes
might imply some further factors than those investigated here.

Besides the perturbations to the $t-J$ Hamiltonian
considered in this paper, it would be interesting to study the effect of 
disorder in HTSC materials which,
according to recent calculations~\cite{sachdev}, 
could be relevant in stabilizing this kind of superstructures.

We thank F. Becca and S. Sorella for important discussions.
M. C. and D. P. acknowledge 
the Agence Nationale de la Recherche (France) for support.
M. R. thanks the Foundation for Polish Science
(FNP) and the Polish Ministry of Science and Education 
under Project No. N202 068 32/1481 for support.


\begin{thebibliography}{99}
\bibitem{note} Indeed, considering that decreasing $J$ along
one direction leads to
smaller $\Delta_{ij}$ for the bonds in that
direction, whereas decreasing
$t$ produces the opposite, we deduce that the kinetic contribution
dominates this competition
(the same effect occurs also in the RVB state with rotational symmetry
breaking).
\bibitem{note2} In presence of lattice distortion, 
different vertical and horizontal bonds
are assumed for the reference homogeneous RVB state.
\end{thebibliography}
\end{document}